\documentclass[12pt]{article}
\usepackage{color}
\usepackage{graphicx}
\usepackage{colortbl}
\usepackage{array}
\usepackage{amssymb}
\usepackage{amsmath}
\usepackage{amsthm}
\usepackage{mathrsfs}
\usepackage{dcolumn}
\usepackage{longtable}
\usepackage{hhline}
\numberwithin{equation}{section}
\allowdisplaybreaks

\title{Three-Parameter Logarithm and Entropy}

\author{\textbf{Cristina B. Corcino}\\{\large\bf Roberto B. Corcino}\\{Research Institute for Computational}\\ {Mathematics and Physics}\\Cebu Normal University\\Cebu City, Philippines \vspace{9pt}}

\begin{document}

\maketitle

\begin{abstract}
A three-parameter logarithmic function is derived using the notion of $q$-analogue and ansatz technique. The derived three-parameter logarithm is shown to be a generalization of the two-parameter logarithmic function of Schw\"ammle and Tsallis as the latter is the limiting function of the former as the added parameter goes to 1. The inverse of the three-parameter logarithm and other important properties are also proved. A three-parameter entropic function is then defined and is shown to be analytic and hence Lesche-stable, concave and convex in some ranges of the parameters.



\end{abstract}

\bigskip

\section{Introduction}
The concept of entropy provides deep insight into the direction of spontaneous change for many everyday phenomena. For example, a block of ice placed on a hot stove surely melts, while the stove grows cooler. Such a process is called irreversible because no slight change will cause the melted water to turn back into ice while the stove grows hotter \cite{Drake}. The concept of entropy was first introduced by German physicist Rudolf Clausius as a precise way of expressing the second law of thermodynamics.

\bigskip

\noindent The Boltzmann equation for entropy is
\begin{equation}
S=k_B\ln \omega,
\end{equation}
where $k_B$ is the Boltzmann constant \cite{Reif} and $\omega$ is the number of different ways or microstates in which the energy of the molecules in a system can be arranged on energy levels \cite{Lambert}. The Boltzmann  entropy plays a crucial role in the foundation of statistical mechanics and other branches of science \cite{151569}.

\bigskip

\noindent The Boltzmann-Gibbs-Shannon entropy \cite{Schwammle-Tsallis, Truffet} is given by
\begin{equation}
S_{BGS} \equiv -k \sum_{i=1}^{\omega} p_i \ln p_i =k \sum_{i=1}^{\omega} p_i \ln \frac{1}{p_i},
\end{equation}
where
\begin{equation}
\sum_{i=1}^{\omega} p_i=1. 
\end{equation}
$S_{BGS}$  is a generalization of the Boltzmann entropy because if  $p_i=\frac{1}{\omega}$, for all $i$,
\begin{equation}
S_{BGS}=k \ln \omega.
\end{equation}
Systems presenting long range interactions and/or long duration memory have been shown not well described by the Boltzmann-Gibbs statistics. Some examples may be found in gravitational systems, L\'evy flights, fractals, turbulence physics and economics. In an attempt to deal with such systems Tsallis \cite{Tsallis1988} postulated a nonextensive entropy which generalizes Boltmann-Gibbs entropy through an entropic index $q$ \cite{9905035}. Another generalization was also suggested by Renyi \cite{Renyi}. Abe \cite{Abe} proposed how to generate entropy functionals.									
\bigskip

\noindent Tsallis $q$-entropy \cite{Tsallis1988} is given by
\begin{equation}
S_q \equiv k \frac{1-\sum_{i=1}^{\omega}p_i^q}{q-1}=k\sum_{i=1}^{\omega} p_i \ln_q \frac{1}{p_i}, 
\end{equation}
where $q\in \mathbb{R}, \sum_{i=1}^{\omega} p_i=1$ and
\begin{equation}
\label{qlog}
\ln_q x \equiv \frac{x^{1-q}-1}{1-q},\;\; (\ln_1 x=\ln x),
\end{equation}
which is referred to as $q$-logarithm. If $p_i=\frac{1}{\omega}$ for all $i$, then 
\begin{equation}
S_q=k \ln_q \omega. 
\end{equation}
The inverse of the $q$-logarithm is the $q$-exponential 
\begin{equation}\label{q-exponential}
e_q^x \equiv [1+(1-q)x]_+^{\frac{1}{1-q}}, \;\; (e_1^x=e^x),
\end{equation}
where $[\cdots]_+$ is zero if its argument is nonpositive. 

\bigskip

\noindent A $q$-sum and $q$-product and their calculus studied in \cite{Borges} were respectively defined as follows (these were also mentioned in \cite{Schwammle-Tsallis}):
\begin{equation}
x \oplus_q y \equiv x+y+(1-q)xy,\;\; (x \oplus_1 y=x+y)
\end{equation}
\begin{equation}
x \otimes_q y \equiv (x^{1-q}+y^{1-q}-1)^{\frac{1}{1-q}}, \;\; (x \otimes_1y=xy). 
\end{equation}
The $q$-logarithm satisfies the following properties:
\begin{equation}
\ln_q(xy)=\ln_q x \oplus_q\ln_q y		
\end{equation}				\begin{equation}
\ln_q(x \otimes_q y)=\ln_q x+\ln_q y.  
\end{equation}
Then a two-parameter logarithm was defined and presented along with a two-parameter entropy in \cite{Schwammle-Tsallis}. It was defined as follows:
\begin{equation}
\label{two}
\ln_{q,q'}x = \frac{1}{1-q'} \left[\exp \left(\frac{1-q'}{1-q} (x^{1-q}-1)\right)-1 \right]. 
\end{equation}
The above doubly deformed logarithm satisfies
\begin{equation}\label{prop1}
\ln_{q,q'}(x \otimes_q y)=\ln_{q,q'}x \oplus_{q'} \ln_{q,q'}y. 
\end{equation}
Properties of the two-parameter logarithm and those of the two-parameter entropy were  proved in \cite{Schwammle-Tsallis}. Probability distribution in the canonical ensemble of the two-parameter entropy was obtained in \cite{Asgarani} while applications were discussed in \cite{Chandra}. 

\bigskip

In section 2 of the present paper, a three-parameter logarithm $\ln_{q,q',r}x$, where  $q,q',r \in \mathbb{R}$,  is derived using $q$-analogues and ansatz technique.  In section 3, the inverse of the three-parameter logarithm is derived and some properties are proved.  A three-parameter entropy and its properties are presented in section 4 and  conclusion is given in section 5.

\section {Three-Parameter Logarithm}

As $x=e^{\ln x}$, a $q$-analogue of $x$ will be defined by
\begin{equation}
[x]_q=e^{\ln_q x},
\end{equation}
where $\ln_q x$ is defined in \eqref{qlog}.
Similarly, the $q'$-analogue of $[x]_q$ is defined by
\begin{equation}\label{qq'analogue}
[x]_{q,q'}=e^{\ln_{q,q'}x}
\end{equation}
where $\ln_{q,q'}x$ is as defined in \eqref{two}, which can be written 
\begin{equation}
\ln_{q,q'}x=\frac{[x]_q^{1-q'}-1}{1-q'}=\frac{\left( e^{\ln_q x}\right)^{1-q'}-1}{1-q'}.
\end{equation}
The three-parameter logarithm is then defined as
\begin{equation}\label{3par}
\ln_{q,q',r}x=\frac{[x]_{q,q'}^{1-r}-1}{1-r}=\frac{\left( e^{\ln_{q,q'} x}\right)^{1-r}-1}{1-r},
\end{equation}
from which
\begin{equation}\label{lnq}
\ln_{q,q',r}x \equiv \frac{1}{1-r}\left\{e^{\left( \frac{1}{1-q'} \left\{e^{(1-q')\ln_q x}-1 \right\} \right)^{1-r}} -1\right\}.
\end{equation}
To obtain similar property as that in \eqref{prop1}, define $x \otimes_{q,q'} y$ as the $q'$-analogue of $x \otimes_q y$. That is,
\begin{equation}\label{prop2}
x \otimes_{q,q'} y\equiv [x \otimes_q y ]_{q'} = \left([x]_{q'}^{1-q}+[y]_{q'}^{1-q}-1 \right)^{\frac{1}{1-q}}. 
\end{equation}
Then, from \eqref{3par} and \eqref{prop2}
\begin{align}
\ln_{q,q'} (x \otimes_{q'} y)&= \frac{[x \otimes_{q'} y]_q^{1-q'}-1}{1-q'} \nonumber\\
&=\frac{\left\{ \left( [x]_q^{1-q'}+[y]_q^{1-q'}-1 \right)^{\frac{1}{1-q'}} \right\}^{1-q'}-1 }{1-q'} \nonumber\\
&=\frac{[x]_q^{1-q'}+[y]_q^{1-q'}-1-1}{1-q'} \nonumber\\
&=\frac{[x]_q^{1-q'}-1}{1-q'}+\frac{[y]_q^{1-q'}-1}{1-q'} \nonumber\\
&=\ln_{q,q'} x+ \ln_{q,q'} y. 
\end{align}
In similar manner and using \eqref{qq'analogue}, 
\begin{align}
\ln_{q,q',r}(x \otimes_{q'} y)&=\frac{[x \otimes_{q'} y]_{q,q'}^{1-r}-1}{1-r} \nonumber\\
&=\frac{\left\{e^{\ln_{q,q'}(x \otimes_{q'} y)} \right\}^{1-r}-1}{1-r}\nonumber\\
&=\frac{\left( e^{\ln_{q,q'}x+\ln_{q,q'}y}\right)^{1-r}-1}{1-r} \nonumber\\
&=\frac{\left(e^{\ln_{q,q' }x} \right)^{1-r} \left( e^{\ln_{q,q'}y}\right)^{1-r}-1 }{1-r} \nonumber\\
&=\frac{\parbox{3in}{$\left\{\left(e^{\ln_{q,q' }x} \right)^{1-r}-1 \right\}+\left\{ \left(e^{\ln_{q,q' }y} \right)^{1-r}-1 \right\} $\\$+\left\{ \left(e^{\ln_{q,q' }x} \right)^{1-r} -1\right\}\left\{ \left(e^{\ln_{q,q' }y} \right)^{1-r} -1\right\}$}}{1-r}. 
\end{align}
Thus, 
\begin{align}
\ln_{q,q',r}(x \otimes_{q'} y)&=\frac{\left(e^{\ln_{q,q' }x} \right)^{1-r}-1 }{1-r}+\frac{\left(e^{\ln_{q,q' }y} \right)^{1-r}-1 }{1-r} \nonumber\\
&\;\;\;\;\;\;\;\;\;\;\;\;\;\;+(1-r) \left[\frac{\left(e^{\ln_{q,q' }x} \right)^{1-r}-1 }{1-r} \right] \left[\frac{\left(e^{\ln_{q,q' }y} \right)^{1-r}-1 }{1-r} \right] \nonumber\\
&=\ln_{q,q',r}x+ \ln_{q,q',r}y+(1-r)[\ln_{q,q',r}x][ \ln_{q,q',r}y]\label{2.11} \\
&=\ln_{q,q',r}x \oplus_r \ln_{q,q',r}y, \label{2.12}  
\end{align}
which is the desired relation analogous to \eqref{prop1}.

\smallskip
One can also derive \eqref{lnq} using ansatz. To do this, let $x=y$ in \eqref{2.12}. Then 
\begin{equation}
\ln_{q,q',r}(x \otimes_{q'} x)=\ln_{q,q',r}x \oplus_r \ln_{q,q',r}x.    
\end{equation}
Taking 
\begin{equation}
\ln_{q,q',r}x=G(\ln_{q,q'}x)=G(z),  
\end{equation}
then
\begin{align}
\ln_{q,q',r}(x \otimes_{q'} x)&=G(
\ln_{q,q'}(x \otimes_{q'} x)) \nonumber\\
&=G(
\ln_{q,q'}x+
\ln_{q,q'}x) \nonumber\\
&=G(2 
\ln_{q,q'}x) \nonumber\\
&=G(2z).  
\end{align}
Thus, from \eqref{2.11} and \eqref{2.12}, 
\begin{align}
G(
2\ln_{q,q'}x)&=
\ln_{q,q',r}x \oplus_r \ln_{q,q',r}x \nonumber\\
&=\ln_{q,q',r}x + \ln_{q,q',r}x + (1-r)(\ln_{q,q',r}x )^2 \nonumber\\
&=2G(\ln_{q,q'}x )+(1-r)[G(\ln_{q,q'}x)]^2 \nonumber\\
G(2z)&=2G(z)+(1-r)[G(z)]^2.  
\end{align}
The ansatz
\begin{equation}\label{ansatz}
G(z)=\frac{1}{1-r}(b^z-1),
\end{equation}
where $z=\ln_{q,q'}x$ will give
\begin{align}
2G(z)+(1-r)[G(z)]^2&=2\cdot \frac{1}{1-r}(b^z-1)+(1-r)\left[\frac{1}{1-r}(b^z-1) \right]^2 \nonumber\\
&=\frac{2}{1-r}(b^z-1)+\frac{(b^z-1)^2}{1-r} \nonumber\\
&=\frac{2b^z-2+b^{2z}-2b^z+1}{1-r} \nonumber\\
&=\frac{b^{2z}-1}{1-r} \nonumber\\
&=G(2z), 
\end{align}
which means that \eqref{ansatz} solves the equation
\begin{equation*}
G(2z)=2G(z)+(1-r)[G(z)]^2. 
\end{equation*}
Thus, 
\begin{equation*}
G(z)=G(\ln_{q,q'}x)=\ln_{q,q',r}x=\frac{1}{1-r}(b^{\ln_{q,q'}x}-1). 
\end{equation*}
Using the property that $\frac{d}{dx} \ln_{q,q',r}x \Big\vert_{x=1}=1$, which is a natural property of a logarithmic function, it is determined that $b=e^{1-r}$. 

\smallskip

\noindent Consequently, 
\begin{equation}
\ln_{q,q',r}x =\frac{1}{1-r}\left(e^{(1-r)\ln_{q,q'}x }-1 \right).  
\end{equation}
Explicitly, 
\begin{equation}\label{explicit formula}
\ln_{q,q',r}x =\frac{1}{1-r}\left(e^{\frac{1-r}{1-q'}\left[\exp \left(\frac{1-q'}{1-q}(x^{1-q}-1) \right)-1 \right]} -1\right),
\end{equation}
which is the same as that in \eqref{lnq}. The preceding equation can be written
\begin{equation}
\ln_{q,q',r}x =\ln_re^{\ln_{q,q'}x}. 
\end{equation}
It can be easily verified that 
\begin{equation}
\lim_{r \to 1} \ln_{q,q',r}x =\ln_{q,q'}x. 
\end{equation}
 Graphs of $\ln_{q,q',r}x$ for $q=q'=r$ are shown in Figure 1 while graphs of $\ln_{q,q',r}x$ with one fixed parameter are shown in Figure 2.

\section{Properties}
In this section the inverse of the three-parameter logarithmic function will be derived. it is also verified that the derivative of this logarithm at $x=1$ is 1 and that the value of the function at $x=1$ is zero. Moreover, it is shown that the following equality holds 
\begin{equation}\label{log2-q}
\ln_{q,q',r}\frac{1}{x}=-\ln_{2-q,2-q',2-r}x.
\end{equation}

It follows from \eqref{3par} that the three-parameter logarithmic function is an increasing function of $x$. Thus, a unique inverse function exists. To find the inverse function let $y=\ln_{q,q',r}(x)$ and solve for $x$. That is,

\begin{equation*}
y=\frac{1}{1-r} \left\{ \exp\left(\frac{1-r}{1-q'}\exp \left(\frac{1-q'}{1-q} (x^{1-q}-1)\right) -1\right)-1 \right\}, 
\end{equation*}
\noindent from which
\begin{equation}
x=\left\{1+\frac{1-q}{1-q'}\ln\left[1+\frac{1-q'}{1-r}
\ln\{1+(1-r)y \} \right] \right\}^{\frac{1}{1-q}}. 
\end{equation}

\noindent Thus,the inverse function is given by 
\begin{align}
e_{q,q',r}^y&=\exp_{q,q',r}y=\left\{1+\frac{1-q}{1-q'}\ln\left[1+\frac{1-q'}{1-r} 
\ln\{1+(1-r)y \} \right] \right\}^{\frac{1}{1-q}}\nonumber\\
&=\left\{1+\frac{1-q}{1-q'}\ln\left[1+(1-q')
\ln\{1+(1-r)y \}^{\frac{1}{1-r}} \right] \right\}^{\frac{1}{1-q}}\nonumber\\
&=\left\{1+\frac{1-q}{1-q'}\ln\left[1+(1-q')
\ln e_r^y \right] \right\}^{\frac{1}{1-q}} \nonumber\\
&=\left\{1+(1-q)\ln\left[1+(1-q')
\ln e_r^y \right]^{\frac{1}{1-q'}} \right\}^{\frac{1}{1-q}}\nonumber\\
&=\left\{1+(1-q)\ln e_{q'}^{\ln e_r^y}
\right\}^{\frac{1}{1-q}}\nonumber\\
&=e_q^{\ln e_{q'}^{\ln e_r^y}}\nonumber\\
&=\exp_q \left\{\ln e_{q'}^{\ln e_r^y} \right\},
\end{align}
where the $q$-exponential $e_q^x$ is defined in \eqref{q-exponential}.  

\bigskip

\noindent To find the derivative, use \eqref{lnq} to obtain
\begin{equation}\label{derivative of 3-parameter log}
\frac{d}{dx}  \ln_{q,q',r} x =x^{-q} \exp \left\{\frac{1-r}{1-q'} (e^{(1-q')\ln_q x}-1)+(1-q')\ln_q x  \right\}.
\end{equation}
\noindent Since $\ln_q 1=0$, it follows that 
\begin{equation}
\frac{d}{dx} \ln_{q,q',r} x \Big\vert_{x=1}=e^0=1.
\end{equation}
\noindent Moreover, 
\begin{equation}
\ln_{q,q',r}1=\frac{1}{1-r}\left\{\exp \left( \frac{1-r}{1-q'}(e^{(1-q')\ln_q 1}-1)\right)-1 \right\}=0.
\end{equation}
From \eqref{derivative of 3-parameter log}, the slope of $\ln_{q,q',r}x$ is positive for all $x>0$. This is also observed in Figures 1 and 2.  

\bigskip
\begin{figure}[t!]
\centerline{\includegraphics[width=15cm]{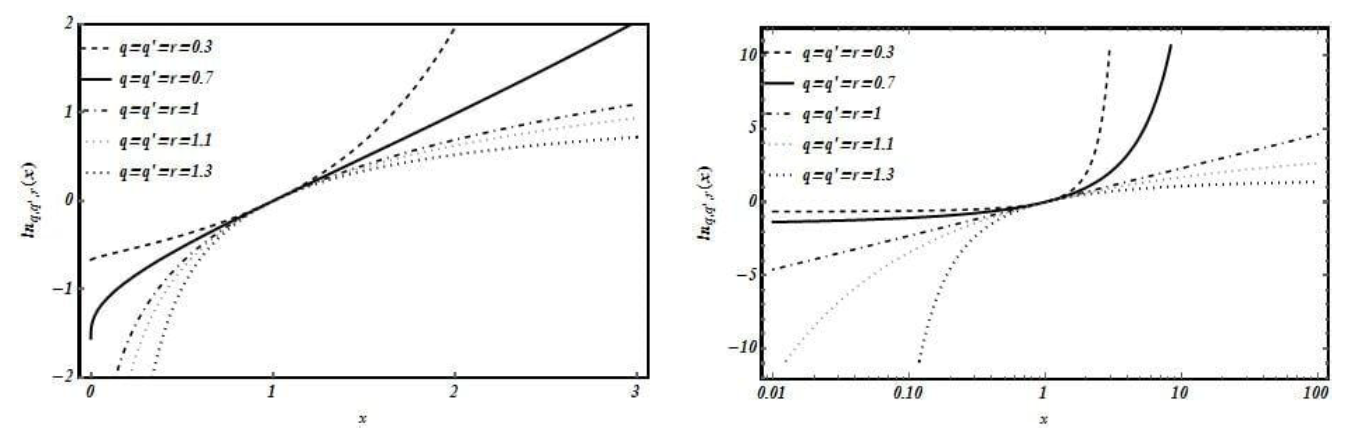}}
\footnotesize{\textbf{Figure 1}}. {\footnotesize{Illustration of the three-parameter logarithm in Eq. \eqref{explicit formula}, setting $q=q'=r$ in linear scales (left) and semi-logarithmic scales (right).}}
\end{figure}

\bigskip

\noindent To prove \eqref{log2-q}, let $q \to 2-q$, $q' \to 2-q'$ and $r \to 2-r$. From \cite{Schwammle-Tsallis},
\begin{equation}\label{prop3}
\ln_{q,q'} \frac{1}{x}=-\ln_{2-q,2-q'} {x},
\end{equation}
then
\begin{align}
\ln_{q,q',r} \frac{1}{x} &=\frac{(e^{\ln_{q,q'} \frac{1}{x}})^{1-r}-1}{1-r} \nonumber\\
&=\frac{(e^{-\ln_{2-q,2-q'}x})^{1-r}-1}{1-r} \nonumber\\
&=\frac{(e^{\ln_{2-q,2-q'}x})^{r-1}-1}{-(r-1)} \nonumber\\
&=\frac{-\{(e^{\ln_{2-q,2-q'}x})^{1-(2-r)}-1\}}{1-(2-r)} \nonumber\\
&=-\ln_{2-q,2-q',2-r}x. 
\end{align}

\section{Three-Parameter Entropy}
A three-parameter generalization of the Boltzmann-Gibbs-Shannon entropy is constructed here and its properties are proved. Based on the three-parameter logarithm the entropic function is defined as follows:
\begin{equation}
S_{q,q^{\prime },r} \equiv k\sum_{i=1}^{\omega}{p_i\ln_{ q,q^{\prime },r} \frac{1}{p_i}}
\end{equation}
If $p_i=\frac{1}{\omega}, \forall i,$
\begin{equation}
S_{q,q^{\prime },r}=k\ln_{q, q^{\prime }, r} {\omega},
\end{equation}
where $w$ is the number of states.\\

\bigskip
\begin{figure}[t!]
\centerline{\includegraphics[width=15cm]{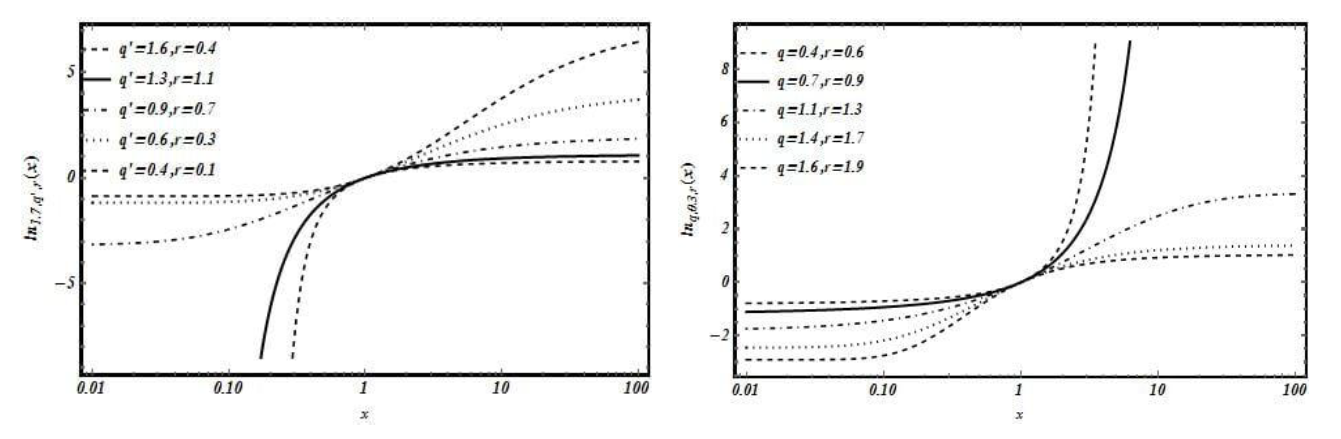}}
\vspace*{1pt}
\end{figure}
\begin{figure}[hbt!]
\centerline{\includegraphics[width=15cm]{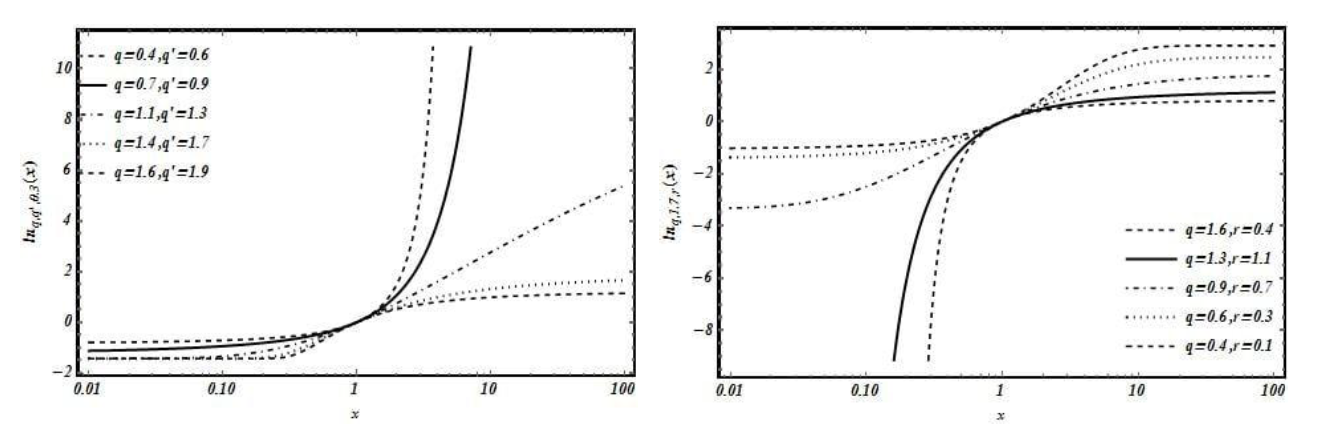}}
\vspace*{1pt}
\end{figure}
\begin{figure}[hbt!]
\centerline{\includegraphics[width=8.5cm]{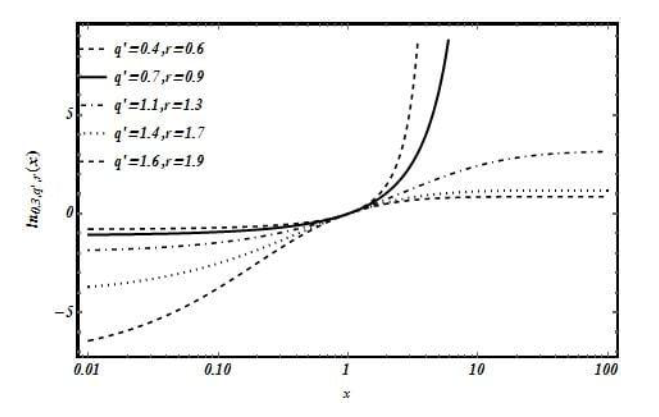}}
\footnotesize{\textbf{Figure 2}}. {\footnotesize{Illustration of the three-parameter logarithm for fixed value of one parameter.}}
\end{figure}

\bigskip
\noindent
\textit{Lesche-stability (or experimental robustness).} The functional form of $\ln_{q, q^{\prime }, r}x$ given in the previous section is analytic in $x$ as $\ln_{q, q^{\prime }}x$ is analytic in $x$. Consequently $S_{q,q^{\prime },r}$ is Lesche-stable.

\bigskip
\noindent
\textit{Expansibility.} An entropic function $S$ satisfies this condition if a zero-probability ($p_i=0$) state does not contribute to the entropy. That is, $S(p_1,p_2, \ldots, p_w,0)= S(p_1,p_2, \ldots, p_w)$ for any distribution $\{p_i\}$. Observe that in the limit $p_i=0$, $\ln_{q,q',r}\frac{1}{p_i}$ is finite if one of $q, q', r$ is greater than 1. Consequently,

\begin{equation}
S_{q,q^{\prime },r}(p_1,p_2, \ldots, p_w,0)= S_{q,q^{\prime },r}(p_1,p_2, \ldots, p_w)
\end{equation}
provided that one of $q,q', r$ is greater than 1.

\bigskip
\noindent
\textit{Concavity.} Concavity of the entropic function $S_q,q',r$ is assured if 
\begin{equation}
\frac{d^2}{dp_i^2}\left(p_i\ln_{q,q',r}\frac{1}{p_i}\right)<0
\end{equation}
in the interval $0\le p_i\le1$. 

\smallskip
By manual calculation (which is a bit tedious) and checked using derivative calculator, 
\begin{align}\label{second derivative}
\frac{d^2}{dp_i^2}\left({p_i}\ln_{q,q',r}\frac{1}{p_i} \right)&=\exp\left\{\frac{1-r}{1-q'}(e^{(1-q')\ln_q\frac{1}{p_i}}-1)\right\}e^{(1-q')\ln_q\frac{1}{p_i}}\;\;\times\nonumber\\
&\;\;\;\;\;\; \left\{-qp_i^{q-2}+(1-q')p_i^{2q-3}+(1-r)p_i^{2q-3}e^{(1-q')\ln_q\frac{1}{p_i}}\right\}.
\end{align}

In the limit $p_i\to1$, the second derivative given in \eqref{second derivative} is less than zero if $q+q'+r>2$. Thus, concavity of $S_{q,q',r}$ is guaranteed if $q+q'+r>2$. In the limit $p_i\to 0$, concavity is guaranteed if $r>1$. If $r<1$, concavity holds if $q>1$.

\bigskip
\noindent
\textit{Convexity.} A twice-differentiable function of a single variable is convex if and only if its second derivative is nonnegative on its entire domain. The analysis on the convexity of $S_{q,q',r}$ is analogous to that of its concavity. In the limit $p_i\to 1$, convexity is guaranteed if $q+q'+r\le 2$. In the limit $p_i\to 0$, convexity is assured if $q,r < 1$. 

\smallskip
Concavity of $S_{q,q'r}$ is illustrated in Figure 3 (A) while convexity is illustrated in Figure 3 (B).

\bigskip
\begin{figure}[t!]
\centerline{\includegraphics[width=15cm]{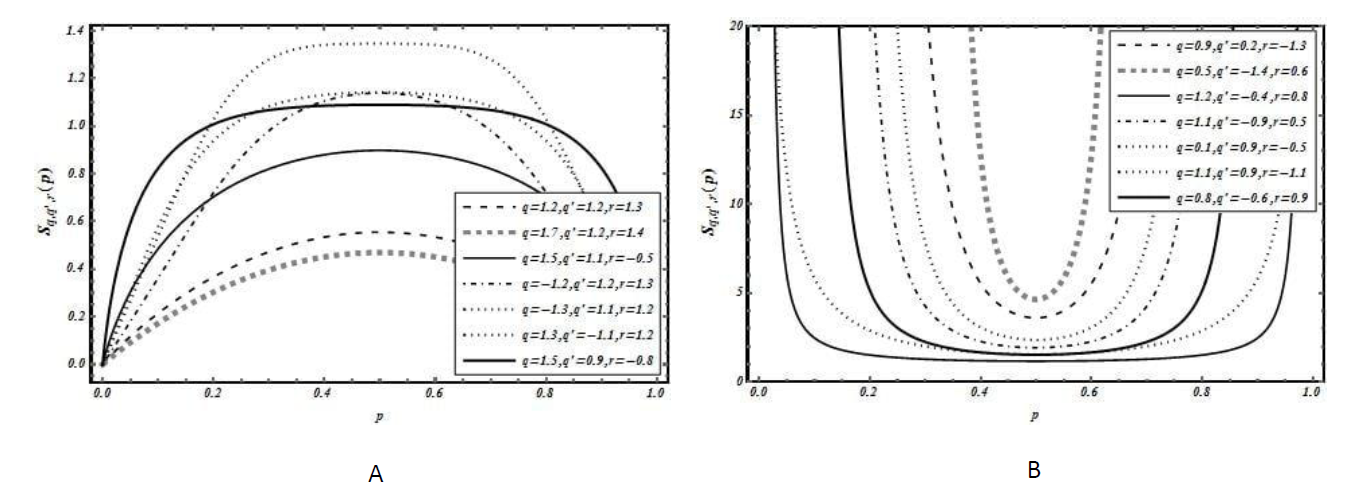}}
{\footnotesize{\textbf{Figure 3}}. \footnotesize{Illustration of the three-parameter entropic function. (A) Concavity. (B) Convexity.}}
\end{figure}

\bigskip
\noindent \textit{Composability.} An entropic function S is said to be composable if for events A and B,
\begin{equation*}
S(A+B)=\Phi(S(A),S(B),indices),
\end{equation*}
where $\Phi$ is some single-valued function \cite{Schwammle-Tsallis}. The Botzmann-Gibbs-Shannon entropy satisfies
\begin{equation*}
S_{BGS}(A+B)=S_{BGS}(A)+S_{BGS}(B),
\end{equation*}
hence it is composable and additive. The one-parameter entropy $S_q$, for $q\neq 1$ is also composable as it satisfies 
\begin{equation}
\frac{S_q^{A+B}}{k}=\frac{S_q^A}{k}\oplus_q\frac{S_q^B}{k}=\frac{S_q(A)}{k}+\frac{S_q(B)}{k}+(1-q)\frac{S_q(A)}{k}\frac{S_q(B)}{k}.
\end{equation}
The two-parameter entropy $S_{q,q'}$ \cite{Schwammle-Tsallis} satisfies, in the microcanonical ensemble (i.e. equal probabilities), that
\begin{equation}\label{star} 
Y(S^{A+B})=Y(S^A)+Y(S^B)+\frac{1-q'}{1-q}Y(S^A)Y(S^B),
\end{equation}
where
\begin{equation}
Y(S)\equiv 1+\frac{1-q}{1-q'}\ln\left[1+(1-q')\frac{S}{k}\right].
\end{equation}
However, this does not hold true for arbitrary distributions $\{p_i\}$, which means $S_{q,q'}$ is not composable in general. For the $3$-parameter entropy $S_{q,q',r}$ a similar property as that of \eqref{star} is obtained as shown below.

\begin{equation}
\ln_{q,q'}(W_AW_b)=\frac{1}{1-q'}\left[e^{(1-q')\ln_q(W_AW_B)}-1\right]=\frac{S_{q,q'}^{A+B}}{k},
\end{equation}
\noindent from which
\begin{equation}\label{S^A}
\frac{S_{q,q',r}^A}{k}=\ln_{q,q',r}w_A=\frac{1}{1-r}\left[e^{(1-r)\ln_{q,q'}w_A}-1\right]
=\frac{1}{1-r}\left[e^{(1-r)\frac{S_{q,q'}^A}{k}}-1\right].
\end{equation}
\noindent Similarly,
\begin{equation}\label{S^B}
\frac{S_{q,q',r}^B}{k}=\ln_{q,q',r}W_B=\frac{1}{1-r}\left[e^{(1-r)\frac{S_{q,q'}^B}{k}}-1\right],
\end{equation}

\begin{equation}\label{S^A+B}
\frac{S_{q,q',r}^{A+B}}{k}=\ln_{q,q',r}w_Aw_B=\frac{1}{1-r}\left[e^{(1-r)\frac{S_{q,q'}^{A+B}}{k}}-1\right]
=\frac{1}{1-r}e^{(1-r)\frac{S_{q,q'}^{A+B}}{k}}-\frac{1}{1-r}.
\end{equation}

\noindent From \eqref{S^A+B},
\begin{equation}\label{S^A+B+1}
\ln\left[(1-r)\frac{S_{q,q',r}^{A+B}}{k}+1\right]=(1-r)\frac{S_{q,q'}^{A+B}}{k}.
\end{equation}
Using the following result in \cite{Schwammle-Tsallis},
\begin{equation}
\frac{S_{q,q'}^{A+B}}{k}=\frac{1}{1-q'}\left\{e^{\frac{1-q'}{1-q}\ln\left[1+(1-q')\frac{S_{q,q'}^A}{k}\right]\ln\left[1+(1-q')\frac{S_{q,q'}^B}{k}\right]\left[1+(1-q')\frac{S_{q,q'}^A}{k}\right]\left[1+(1-q')\frac{S_{q,q'}^B}{k}\right]}-1\right\}
\end{equation}
\eqref{S^A+B+1} becomes
\begin{multline*}
\ln\left[1+(1-r)\frac{S_{q,q',r}^{A+B}}{k}\right]=\frac{1-r}{1-q'}\left\{e^{\frac{1-q'}{1-q}\ln\left[1+\frac{1-q'}{1-r}\ln\left[1+(1-r)\frac{S_{q,q',r}^A}{k}\right]\right]\cdot \ln\left[1+\frac{1-q'}{1-r}\ln\left[1+(1-r)\frac{S_{q,q',r}^B}{k}\right]\right]}\right.\\
\left.\times \left[1+\frac{1-q'}{1-r}\ln\left[1+(1-r)\frac{S_{q,q',r}^A}{k}\right]\right]\right.\\
\left.\times \left[1+\frac{1-q'}{1-r}\ln\left[1+(1-r)\frac{S_{q,q',r}^B}{k}\right]\right]-1\right\}.
\end{multline*}
Let 
\begin{equation}
U(S)=\ln\left[1+\frac{1-q'}{1-r}\ln\left[1+(1-r)\frac{S}{k}\right]\right].
\end{equation}
\noindent Then
\begin{multline*}
1+\frac{1-q'}{1-r}\ln\left[1+(1-r)\frac{S_{q,q',r}^{A+B}}{k}\right]=e^{\frac{1-q'}{1-q}U(S^A)\cdot U(S^B)} \\
\times \left[1+\frac{1-q'}{1-r}\ln\left[1+(1-r)\frac{S_{q,q',r}^A}{k}\right]\right] \\
\times \left[1+\frac{1-q'}{1-r}\ln\left[1+(1-r)\frac{S_{q,q',r}^B}{k}\right]\right].
\end{multline*}
\noindent Consequently,
\begin{multline*}
\ln\left[1+\frac{1-q'}{1-r}\ln\left[1+(1-r)\frac{S_{q,q',r}^{A+B}}{k}\right]\right]=\frac{1-q'}{1-q}U(S^A)\cdot U(S^B) \\
+\ln\left[1+\frac{1-q'}{1-r}\ln\left[1+(1-r)\frac{S_{q,q',r}^A}{k}\right]\right] \\
+\ln\left[1+\frac{1-q'}{1-r}\ln\left[1+(1-r)\frac{S_{q,q',r}^B}{k}\right]\right],
\end{multline*}
\noindent which can be written
\begin{equation}
U(S^{A+B})=U(S^A)+U(S^B)+\frac{1-q'}{1-q}U(S^A)U(S^B).
\end{equation}

\noindent In view of the noncomposability of the 2-parameter entropy, $S_{q,q',r}$ is also non-composable.

\section{Conclusion}
It is shown that the two-parameter logarithm of Schwammle and Tsallis \cite{Schwammle-Tsallis} can be generalized to three-parameter logarithm using q-analogues. Consequently, a three-parameter entropic function is defined and its properties are proved. It will be interesting to study applicability of the three-parameter entropy to adiabatic ensembles \cite{Chandra} and other ensembles \cite{Roussel} and how these applications relate to generalized Lambert W function.

\section*{Acknowledgment}
This research is funded by Cebu Normal University (CNU) and the Commission  on Higher Education - Grants-in-Aid (CHED-GIA) for Research.

\section*{Data Availability Statement}
The computer programs and articles used to generate the graphs and support the findings of this study are available from the corresponding author upon request.

\end{document}